# Valley-polarized Exitonic Mott Insulator in WS$_2$/WSe$_2$ Moiré Superlattice

Zhen Lian[1#], Yuze Meng[1#], Lei Ma[1#], Indrajit Maity[2#], Li Yan[1], Qiran Wu[3], Xiong Huang[3,4], Dongxue Chen[1], Xiaotong Chen[1], Xinyue Chen[1], Mark Blei[5], Takashi Taniguchi[6], Kenji Watanabe[7], Sefaattin Tongay[5], Johannes Lischner[2], Yong-Tao Cui[3*], Su-Fei Shi[1,8*]

1. Department of Chemical and Biological Engineering, Rensselaer Polytechnic Institute, Troy, NY 12180, USA
2. Departments of Materials and the Thomas Young Centre for Theory and Simulation of Materials, Imperial College London, United Kingdom.
3. Department of Physics and Astronomy, University of California, Riverside, California, 92521, USA
4. Department of Materials Science and Engineering, University of California, Riverside, California, 92521, USA
5. Materials Science and Engineering Program, School for Engineering of Matter, Transport and Energy, Arizona State University, Tempe, AZ 85287, USA
6. International Center for Materials Nanoarchitectonics, National Institute for Materials Science,1-1 Namiki, Tsukuba 305-0044, Japan
7. Research Center for Functional Materials, National Institute for Materials Science, 1-1 Namiki, Tsukuba 305-0044, Japan
8. Department of Electrical, Computer & Systems Engineering, Rensselaer Polytechnic Institute, Troy, NY 12180, USA

[#] These authors contributed equally to this work
[*] Corresponding authors: shis2@rpi.edu, yongtao.cui@ucr.edu

## Abstract

Strongly enhanced electron-electron interaction in semiconducting moiré superlattices formed by transition metal dichalcogenides (TMDCs) heterobilayers has led to a plethora of intriguing fermionic correlated states[1–7]. Meanwhile, interlayer excitons in a type-II aligned TMDC heterobilayer moiré superlattice, with electrons and holes separated in different layers[8,9], inherit this enhanced interaction and strongly interact with each other, promising for realizing tunable correlated bosonic quasiparticles[10,11] with valley degree of freedom. We employ photoluminescence spectroscopy to investigate the strong repulsion between interlayer excitons and correlated electrons in a WS$_2$/WSe$_2$ moiré superlattice and combine with theoretical calculations to reveal the spatial extent of interlayer excitons and the band hierarchy of correlated states. We further find that an excitonic Mott insulator state emerges when one interlayer exciton occupies one moiré cell, evidenced by emerging photoluminescence peaks under increased optical excitation power. Double occupancy of excitons in one unit cell requires overcoming the energy cost of exciton-exciton repulsion of about 30-40 meV, depending on the stacking configuration of the WS$_2$/WSe$_2$ heterobilayer. Further, the valley polarization of the excitonic Mott insulator state is enhanced by nearly one order of magnitude. Our study demonstrates the WS$_2$/WSe$_2$ moiré superlattice as a promising platform for engineering and exploring new correlated states of fermion, bosons, and a mixture of both[12].

## Main Text

Semiconducting moiré superlattices of transition metal dichalcogenide (TMDC) monolayers, especially that formed by the $WS_2/WSe_2$ heterobilayer, have been shown to exhibit strong electron correlation due to the formation of flat moiré miniband and strong Coulomb interaction[1–5]. In two-dimensional (2D) systems, in general, the Coulomb interaction is enhanced because of reduced dielectric screening. In a TMDC moiré superlattice, the significantly larger effective mass of flat moiré miniband[13] further reduces the kinetic energy compared to that in TMDC monolayers. As a result, the degree of electron-electron interaction is strongly enhanced. Exciting electronic correlated states with high transition temperatures have been demonstrated and investigated. In particular, the strong electron interaction can result in the spatial ordering of charge carriers in the moiré superlattice, forming Mott insulator and generalized Wigner crystal states[1–4,14,15]. Meanwhile, the strong Coulomb interaction in TMDC semiconductors also leads to strongly bound electron-hole pairs, i.e., robust excitons with valley degree of freedom[16–18]. It is thus intriguing to investigate whether these optically excited excitons would form strongly correlated bosonic states in a TMDC moiré superlattice due to reduced kinetic energy, which will allow the exploration of bosonic correlated physics in a highly tunable two-dimensional condensed matter system.

In this work, we turn to the archetypal $WS_2/WSe_2$ moiré superlattice whose type-II band alignment enables interlayer excitons (IX) with holes residing in the $WSe_2$ layer and electrons in the $WS_2$ layer. The permanent dipole moment stemming from the charge separation gives rise to strong repulsion between IXs, rendering this system a promising candidate for realizing the excitonic Mott insulator, a strongly correlated state described by the bosonic Hubbard model[19]. We first use photoluminescence (PL) spectroscopy to show the strong repulsion between electrons and IX at correlated insulating states, which we combine with first-principles calculations to reveal the more extended nature of the IX in the H-stacked (60° twisted) $WS_2/WSe_2$ moiré superlattice than the R-stacked (0° twisted) configuration. We further show that adding excitons to a charge-neutral moiré superlattice with a density high enough to excite double IX occupation in one moiré unit cell will result in additional high energy PL emissions due to IX-IX repulsion, which is about 44 meV for the R-stacking and about 32 meV for the H-stacking. Interestingly, the valley polarization of the IX is strongly enhanced by nearly one order of magnitude when the excitonic Mott insulator state emerges, inspiring future exploration of valley physics in this strongly correlated bosonic system.

We construct angle-aligned $WS_2/WSe_2$ moiré bilayers using the same dry pick-up method as described in our previous works[3,20–22] and fabricate devices with dual-gate geometry as schematically shown in Fig. 1a. The application of top and back gate voltages allows us to control the doping and electric field independently. The doping-dependent PL spectra for both R- and H-stacked moiré bilayers can be found in Figs. 1c (device R1) and 1d (device H1). The enhanced PL intensity at characteristic carrier densities marks the formation of correlated insulating states at filling factors n=1, 2, 3, i.e., one, two, or three electrons per moiré unit cell, and also on the hole side at n= -1 and -2. Correlated states at fractional fillings of n=1/3, 2/3 are also visible, corresponding to generalized

Wigner crystal states studied previously[1,3]. The zoom-in PL spectra showing fractional filling more clearly for the devices shown in Fig. 1c and Fig. 1d are shown in Extended Data Fig. 1. The PL spectra of another H-stacked device (H5) showing more fractional fillings can be found in Extended Data Fig. 2.

The most pronounced features in Figs. 1c and 1d are the abrupt blueshifts of IX resonance energy (PL peak) at Mott insulating states of n=-1 and 1. The blueshift for the R-stacked device (R1) is ~17 meV and ~9 meV for n=1 and -1; while the blueshift for the H-stacked device (H1) is ~41 meV and ~19meV for n=1 and -1, respectively. We first note that the magnitude of the blueshift at n=±1 is smaller for the R stacking than the H stacking. Secondly, the blueshift at n=-1 is smaller than that at n=1 in both stackings.

The blueshifts of IX PL at Mott insulating states n=±1 originate from the electron-exciton repulsion. For example, at the Mott insulator state at n=1 in which each moiré unit cell is occupied by one electron, adding an IX to any moiré unit cell requires an additional energy cost on the scale of the Mott gap ($U_{e-e}$) due to the close proximity of the correlated electron and the constituent electron of the IX in the same $WS_2$ layer. This energy cost will be offset by the Coulomb attraction between the correlated electron in the $WS_2$ layer and the constituent hole of the IX in the $WSe_2$ layer ($V_{e-h}$), which is smaller than $U_{e-e}$ due to layer separation. This interpretation of the blueshift, combined with theoretical insights gained from first-principles calculations, can explain the two major observations described previously.

The first-principles calculations of band structures for both R- and H-stacked $WS_2/WSe_2$ moiré heterobilayers are shown in Figs. 2a and 2b, respectively. For both R- and H-stackings, there is one flat valence band (V1) isolated from other valence bands with an energy separation $\Delta_h$. There are also two flat conduction bands (C1 and C2) with an energy separation, $\Delta_e$, between them. At K and K' valleys, the spin is polarized in the valence band, while C1 and C2 possess different spins. The values of $\Delta_h$ for R- and H-stackings are different, 26 meV and 35 meV, respectively. On the other hand, and $\Delta_e$ is very similar for both R- and H- stackings, 31 and 32 meV, respectively.

It is evident that R- and H-stackings differ significantly in the electronic wavefunction distribution for each of the flat bands (see Fig. 2c and Fig. 2d). In either R or H-stacked configuration, there are three high symmetry points that preserve the three-fold rotational symmetry (insets of Figs. 2c, d), which we label with their local stacking configurations. The moiré potential landscape varies at different moiré sites. In the R-stacking, the holes from V1 and electrons from both C1 and C2 are all localized in the $B^{Se/W}$ site. In contrast, in the H-stacking, the holes from V1 are localized in the $B^{W/W}$ site, and the electrons from the C1 and C2 are localized at the AA' site (2H). As the IX ground state consists of an electron from C1 and a hole from V1, the IX will have the electron and hole at the same moiré site in the R-stacking but at different moiré sites in the H-stacking. As a result, the IX will be more extended in the H-stacking, consistent with the shallower moiré potential confinement of excitons in the H-stacking from our calculation (details in Supplementary Information Section 8) as well as a recent report[23]. The extended nature of IX in the H-stacking is also consistent with our experimental observations in Figs. 1e, f: in the H-stacked device, the IX PL peak redshifts as soon as electrons or holes are electrostatically introduced, likely due to the sensitivity of the extended IX to the dielectric screening; in contrast, the IX PL barely changes between n=-1 and 1 in the R-stacked device.

We now examine the electron-IX interaction based on these understandings. At n = 1, the lower Hubbard band (LHB) of C1 is fully occupied with electrostatically doped electrons. Therefore, the IX consists of an electron in the upper Hubbard band (UHB) of C1 and a hole in the flatband V1, as illustrated in Fig. 2e. The electrons in the LHB of C1 will interact with the IX electron-hole pair and contribute both a repulsion term, $U_{e-e}$, and an attraction term, $V_{e-h}$. The total electron-IX repulsion energy is thus $U_{e-IX}(n=1) \approx U_{e-e} - V_{e-h}$, which corresponds to the energy blueshift observed in our experiment. The scenario at n=-1 is similar (see Fig. 2f), except that the repulsion is between IX hole and the holes in the UHB of V1 with a repulsion term $U_{h-h}$ and an attraction term $V_{e-h}$, then we can write $U_{e-IX}(n=-1) \approx U_{h-h} - V_{e-h}$.

The first major experimental observation can be understood with the extended IX in the H-stacking, which suggests a smaller $V_{e-h}$. As a result, we expect the blueshift to be larger in H-stacked devices, which explains our first major observation in Figs. 1e,f that the blueshifts in H stacked device are larger at n=±1. Secondly, we have estimated the onsite repulsion energies ($U_{e-e}$, $U_{h-h}$) for different valence and conduction bands using the localization of the wavefunctions (details in Supplementary Information Section 9). The electron-electron repulsion ($U_{e-e}$) is larger than hole-hole repulsion ($U_{h-h}$) in the same stack configuration, which well explains our second major observation of the larger blueshift observed for n=1 than n=-1 in Figs. 1e,f.

In Figs, 1e,f, we also observe that the blueshift is about 0 meV at both n = ± 2 for the R-stacked device, while the blueshift is ~17 meV and ~15 meV for the H-stacked device at n=+2 and -2, respectively. We notice that the blueshift is generally smaller at n = ± 2 compared to those at n = ± 1 for both R- and H-stacked devices, which coincide with previous observations that the n=±1 states have a higher transition temperature than n=±2 hence a larger energy gap at n=±1[2–4]. We also observe an additional PL peak around 1.39 eV on the hole side in R-stacked devices, potentially due to the attractive interlayer trion with a binding energy of about 8 meV, which is not the focus of the current work and will be explored later.

Understanding the blueshift at the n=±2 states is more complicated, as it involves the nature of the insulating states, the bandgaps ($\Delta_e$ and $\Delta_h$), and the repulsion energy involving C2 and V2. In Supplementary Information Section 10, we provide a simple model to attempt to explain the observations, yet a complete understanding requires future work to refine the model and calculations. We emphasize here that our experimental observations of n=±2 states, shown in Figs. 1e,f, are reproduced in over six R-stacked and six H-stacked devices, providing valuable guidance for further theoretical understandings.

The strong Coulomb interaction in the WS$_2$/WSe$_2$ moiré superlattice enhances not only the electron-IX interaction but also the IX-IX interaction, making it possible to realize the bosonic Mott state composed of IXs. Indeed, as we increase the optical excitation power, new PL peaks emerge at higher energies, labeled as $X_i^2$ and $X_i^3$ in Figs. 3a, b. The PL peak $X_i^2$ has an onset excitation power of 1-25 µW in most devices, and we have reproduced $X_i^2$ in six R-stacked devices and six H-stacked devices. The detailed excitation power dependence of the doping-dependent PL spectra of six typical devices are shown in Extended Data Fig. 3. The PL peak $X_i^3$ occurs at a much higher excitation power, typically 100s' µW (Figs. 3a). We have only observed it in four R-stacked devices, not in any of the H-stacked devices. We rule out the possibility of these new PL peaks being the excited states of IX from other moiré sites or conduction bands, as those should exhibit an opposite valley polarization to the ground state IX, which contradicts our observations in Fig. 4[24]. Further, the fact that the higher energy PL peaks are absent for n>1 in the H-stacking also rules out those possibilities.

We attribute $X_i^2$ to the blue-shifted IX PL in the presence of a bosonic Mott insulator of IXs (Fig. 3c)[25], which emerges when each moiré unit cell is filled with one IX. Any additional IX will then create a double occupancy in one moiré unit cell. Hence, the IX-IX repulsion increases the IX energy. This state is simply a bosonic analog of an electron Mott insulator. To examine this picture, we measure the power dependence of the PL spectra in the R-stacked device R1 (see Fig. 3d). The extracted linewidth of $X_i^2$ decreases when the excitation power increases from 1 to 20 µW (blue shaded area in Fig. 3e). This behavior is surprising because an increased IX-IX interaction at higher IX densities is expected in general to shorten the IX lifetime, which would broaden the linewidth. On the other hand, it can be well explained if we consider the formation of a bosonic Mott insulator of IXs in which IXs are localized. The excitation power at which the linewidth starts to decrease also coincides with the onset of the $X_i^2$ as marked by the white arrow in Fig. 3d. Furthermore, we estimate the order of magnitude of excitation power needed to generate an IX density of one IX per moiré unit cell to be about 10 µW (see Supplementary Section 5), which is consistent with the typical onset excitation power in the range of 1-25 µW observed in six R-stacked devices and five H-stacked devices. A more detailed analysis of the excitation power dependence can be found in Supplementary Information Section 5.

The IX-IX repulsion energy (U$_{ex-ex}$), which we estimate from the energy difference between $X_i^2$ and $X_i$, is about 44 meV for the R-stacked device R1 and about 32 meV for the H-stacked device H1. The reduced IX-IX repulsion is consistent with the extended nature of the IX wavefunction in the H-stacking, which suggests a shallower energy trap for IX[26]. For the same reason, the H-stacked device cannot host three excitons in one moiré unit cell, but the R-stacked one can ($X_i^3$), albeit with a reduced IX-IX repulsion, about 20 meV estimated from the difference between $X_i^3$ and $X_i^2$. The apparent asymmetry of $X_i^2$ (and $X_i^3$ in R-stacking) as a function of the doping remains an intriguing subject to be explored in the future. It is worth noting here that the observed U$_{ex-ex}$ is more than one order of magnitude larger than that in the double quantum wells made of GaAs/AlGaAs[27,28], thus enabling an excitonic Mott state at a much higher temperature.

Finally, we have performed helicity-resolved PL spectroscopy measurements in both R- and H-stacked devices (R1 and H1) using continuous wave (CW) laser excitation centered at 1.70 eV, resonantly exciting the moiré A exciton of WSe$_2$ (details in the Methods section). The extracted PL circular polarization, $\rho$, is plotted as a function of doping, as shown in Fig. 4. Under a low excitation power of 2 μW (Fig. 4a), for the R-stacked device R1, the PL is co-circularly polarized with respect to the excitation in both charge neutral and electron-doped regions, while the PL is cross-polarized in the highly p-doped region. However, for the H-stacked device H1 (Fig. 4c), $\rho$ is close to zero near the charge-neutral region but cross-polarized in both highly n- and p-doped regions. The circular polarization of the R- and H-stacked WS$_2$/WSe$_2$ moiré superlattice can be well understood by considering the optical selection [29,30], the long valley lifetime of holes, and the efficient intervalley scattering of electrons[31], owing to the vastly different spin-orbit coupling magnitude in the conduction and valence bands. Detailed explanations can be found in Supplementary Information Section 11.

Here we note that above n=2 the circular polarization of the emerging IX PL branch near 1.46 eV in the R-stacked sample reverses sign, suggesting that the added interlayer exciton will have its electron occupying the second conduction band (the LHB of C2 to be precise, with an opposite spin as opposed to C1 (detailed discussion in Supplementary Information Section 11).

As we increase the excitation power, there are two major changes to the circular polarization spectra in the charge-neutral region of device R1 (Fig. 4a,b). First, $\rho$ of the interlayer exciton ground state ($X_i$) increases significantly. Second, the second PL peak ($X_i^2$) also exhibit a finite valley polarization that has the same sign as $X_i$. The latter rules out the possibility of $X_i^2$ being excitons from another moiré site or conduction band, as those will switch the sign of $\rho$. The enhanced circular polarization is striking, which increases from 7% (Fig. 4a) to 50 % (Fig. 4b) in the charge-neutral region n=0, an increase by nearly one order of magnitude. The presence of $X_i^2$ indicates the formation of the bosonic Mott insulator state. The enhancement of PL valley polarization of $X_i$, therefore, is strongly tied to the strong exciton correlation. The significantly enhanced valley polarization could possibly arise from the suppressed valley scattering of IXs at the bosonic Mott insulator state or even optical pumping induced ferromagnetism[32,33], while its exact nature needs to be further explored. The valley-polarized excitons usher new venues for exploring correlated excitons that do not exist in other systems and inspire future exploration of valley physics in strongly correlated bosonic systems.

**Methods**

**Device Fabrication**

We have employed a dry pick-up method, as reported in our earlier work[22] to fabricate angle-aligned WS₂/WSe₂ heterobilayers. For the device shown in the main text, during the pick-up process, we tear one WS₂ monolayer into two pieces, rotate one piece by 180° and align both pieces with one single WSe₂ monolayer to get both R-stacked and H-stacked devices on one chip. The fabricated devices were then annealed at 250 °C in a vacuum for 500 min.

**Optical Characterizations**

The optical measurements were performed under a home-built confocal microscope, with the sample cooled down by either a liquid helium-flow-controlled optical cryostat (Janis) or a cryogen-free optical cryostat (Montana Instruments). The excitation laser was focused on the sample with a beam spot diameter of about 2 µm, and the optical signals were collected by a spectrometer (Princeton Instruments). The PL measurements in Figs. 1, 3a, 3b, and Extended Data Figs. 1, 3 were performed with a CW laser centered at 1.96 eV. All other PL measurements were performed with a CW laser centered at 1.70 eV. To perform helicity-resolved measurements, a quarter waveplate was placed before the objective to convert the excitation laser to σ⁺ polarized light and circularly polarized PL emission to linearly polarized light. The σ⁺ and σ⁻ PL emission was analyzed using a half waveplate and a polarizer. The doping-dependent PL spectra in the right ($\sigma^+$) and left circularly ($\sigma^-$) polarized channels can be found in Extended Data Fig. 4-6. The extracted PL circular polarization expressed as $\rho = \frac{I^+ - I^-}{I^+ + I^-}$, where $I^+$ and $I^-$ denote the intensity of $\sigma^+$ and $\sigma^-$ PL emission. The reflectance contrast measurements were performed with a super-continuum laser (YSL Photonics). The polarized SHG measurements were performed with a pulsed laser excitation centered at 800 nm (Ti: Sapphire; Coherent Chameleon) with a repetition rate of 80 MHz and a power of 100 mW. The crystal axes of the sample were fixed. A half waveplate is placed between the beam splitter and the objective to change the polarization angle of both the excitation laser and the SHG signal. The SHG signal was analyzed using a polarizer. The details of time-resolved photoluminescence measurements and exciton density estimate can be found in Supplementary Sections 4 and 5, respectively.

**Electronic Structure Calculations**

The twisted WS₂/WSe₂ heterobilayer structures are generated using the TWISTER package[34]. Structural relaxations are performed using the LAMMPS (Large-scale Atomic/Molecular Massively Parallel Simulator) package[35,36]. The electronic structure calculations are performed using the SIESTA (Spanish Initiative for Electronic Simulations with Thousands of Atoms) package[37] based on density functional theory (DFT)[38]. See Supplementary Section 7 for details.

**Data Availability**

Source data are available for this paper. All other data that support the plots within this paper and other findings of this study are available from the corresponding author upon reasonable request.

**Code Availability**

The twisted bilayer structure construction, atomic relaxations, and electronic band structure calculations presented in the paper were carried out using publicly available codes. Our findings can be fully reproduced by the use of these codes and by following the procedure outlined in the paper.

**Acknowledgments**

We thank Prof. Chenhao Jin for the helpful discussions. Z. Lian and S.-F.S. acknowledge support from NYSTAR through Focus Center-NY–RPI Contract C150117. The device fabrication was supported by the Micro and Nanofabrication Clean Room (MNCR) at Rensselaer Polytechnic Institute (RPI). S.-F. S. also acknowledges the support from NSF (Career Grant DMR-1945420, DMR-2104902, and ECCS-2139692). XH and Y.-T.C. acknowledge support from NSF under award DMR-2104805. The optical spectroscopy measurements are also supported by a DURIP award through Grant FA9550-20-1-0179. MI acknowledges funding from the European Union's Horizon 2020 research and innovation program under the Marie Skłodowska-Curie Grant agreement No. 101028468. This work used the ARCHER2 UK National Supercomputing Service (https://www.archer2.ac.uk). ST acknowledges support from DOE-SC0020653, Applied Materials Inc., NSF CMMI 1825594, NSF DMR-1955889, NSF CMMI-1933214, NSF DMR-1904716, NSF 1935994, and NSF ECCS 2052527 and DMR 2111812. KW and TT acknowledge support from the Elemental Strategy Initiative conducted by the MEXT, Japan, Grant Number JPMXP0112101001 and JSPS KAKENHI, Grant Numbers 19H05790 and JP20H00354.

**Author Contributions**

S.-F. S. conceived the project. Y. M., Z. L., and DC fabricated heterostructure devices. Z. L., LM, L. Y., and Y. M. performed the optical spectroscopy measurements. MB and ST grew the TMDC crystals. TT and KW grew the BN crystals. IM and JL performed the DFT calculations. S.-F. S, Y.-T. C., Z. L., Y. M., Xiaotong C., and Xinyue C. analyzed the data. S.-F. S. and Y.-T. C. wrote the manuscript with input from all authors.

**Competing Interests**

The authors declare no competing interests.

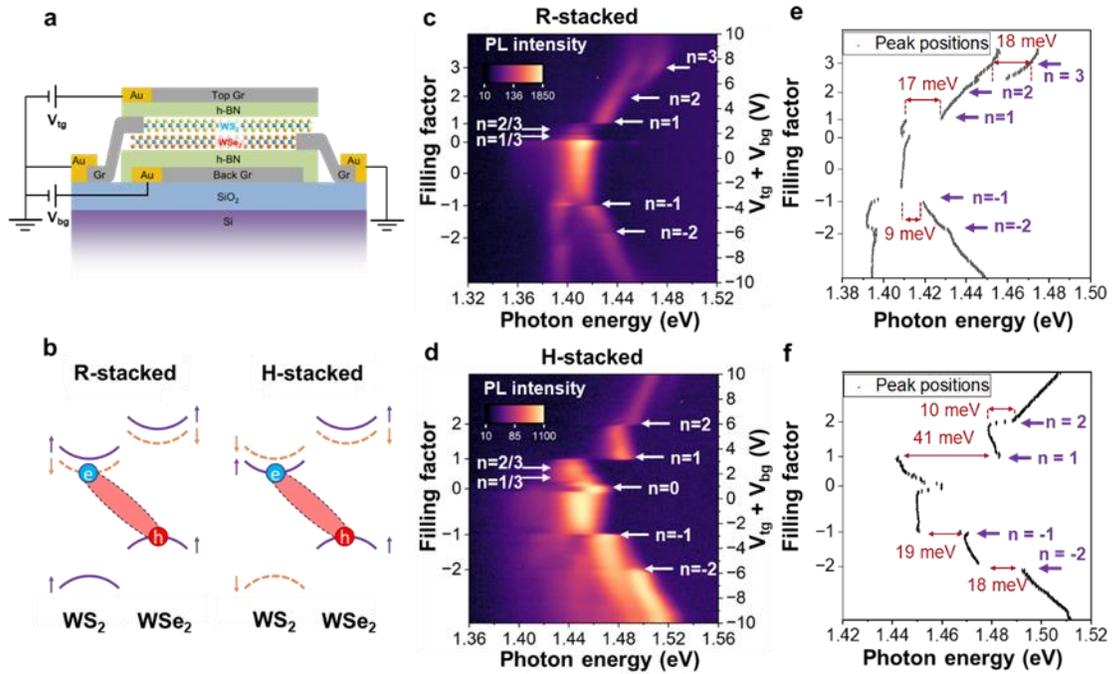

**Fig.1 | Doping-dependent PL spectra of interlayer excitons in WS$_2$/WSe$_2$ moiré superlattice.** (a) Schematic of the WS$_2$/WSe$_2$ device. (b) Schematic of the type II alignment for both R-stacked and H-stacked configurations. (c) and (d) are doping-dependent PL spectra of the R-and H-stacked device, respectively. (e) and (f) are PL peaks extracted from (c) and (d) by fitting the PL spectra with Lorentzian peaks. The error bars indicate the standard deviations of the fitting. The PL spectra were taken with a CW laser excitation centered at 1.96 eV. The excitation power is 5 µW in (c) and 25 µW in (d), corresponding to an average exciton density around 2.9×10$^{11}$ cm$^{-2}$ and 1.45×10$^{12}$ cm$^{-2}$, respectively. All spectra were taken at a temperature of 4.2 K.

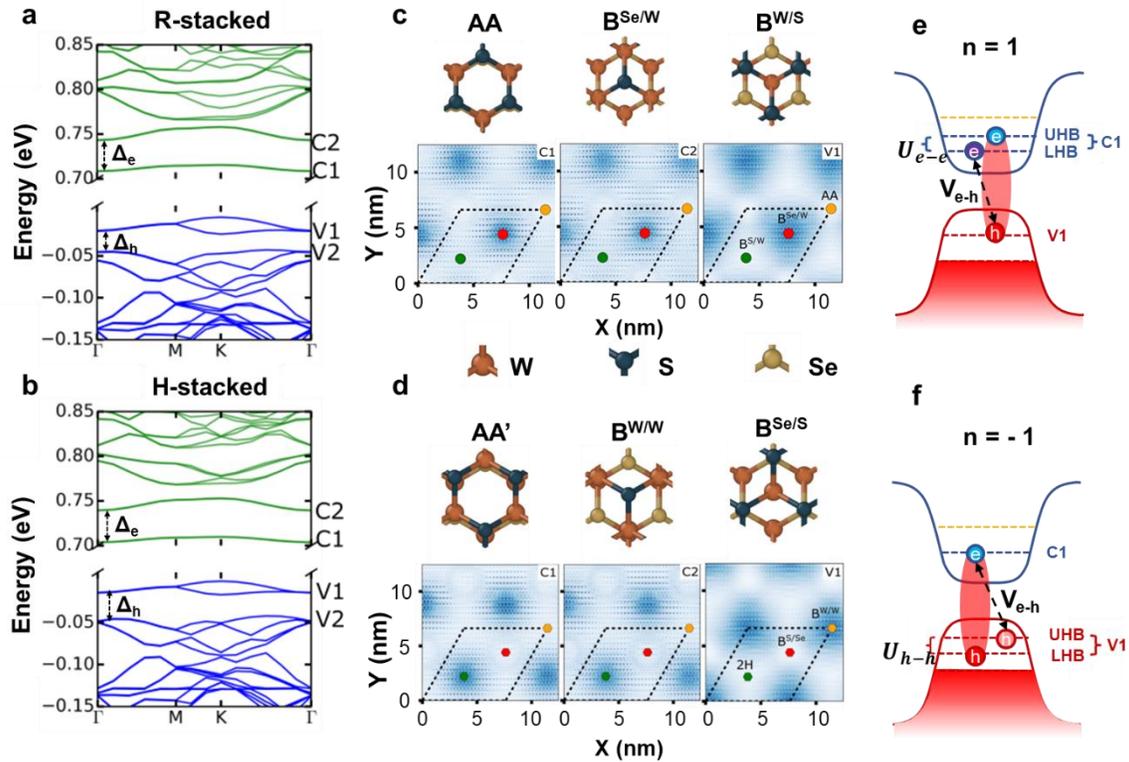

**Fig.2 | Bandstructure of the WS$_2$/WSe$_2$ moiré superlattice.** "(a) and (b) are first-principles calculations of the bandstructure of R- and H-stacked WS$_2$/WSe$_2$ moiré superlattice. The bottom panel of (c) and (d) shows the electron wave function distribution of two conduction bands (C1 and C2) and one valence band (V1) in the moiré unit cell for R- and H-stacked configurations, respectively. The unit cell is marked with dashed lines and the high-symmetry stackings associated with the moiré are marked with circles. The top panel of (c) and (d) shows the atomic structure of the associated high symmetry stackings (e) and (f) are the schematics of adding one interlayer exciton to the correlated insulating state at filling n = 1, and -1.

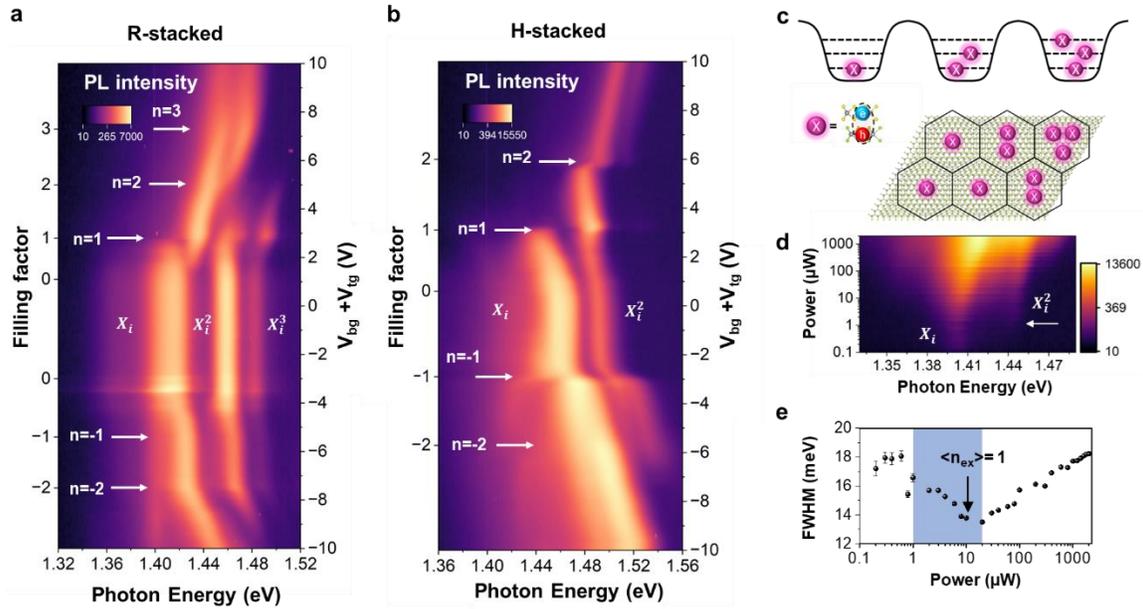

**Fig.3 | Excitonic Mott state formation with increased exciton density.** (a) and (b) are doping-dependent PL spectra with an increased optical excitation power of 300 μW (corresponds to an average exciton density around $1.7\times10^{13}$ cm$^{-2}$), with the CW excitation photon energy centered at 1.96 eV. (c) is a schematic of the excitonic Mott insulator with some moiré unit cells having double or triple exciton occupancy. (d) is the excitation power dependence of the normalized PL spectra of R-stacked device R1, with a CW excitation photon energy centered at 1.70 eV (e) is the full width at half maximum (FWHM) of the fitted Lorentzian peaks extracted from (d), plotted as a function of the optical excitation power. The error bars indicate the standard deviations of the fitting. The blue-shaded region highlights the FWHM decreases with the increasing excitation power. All the data were taken at a temperature of 4.2 – 5.0 K.

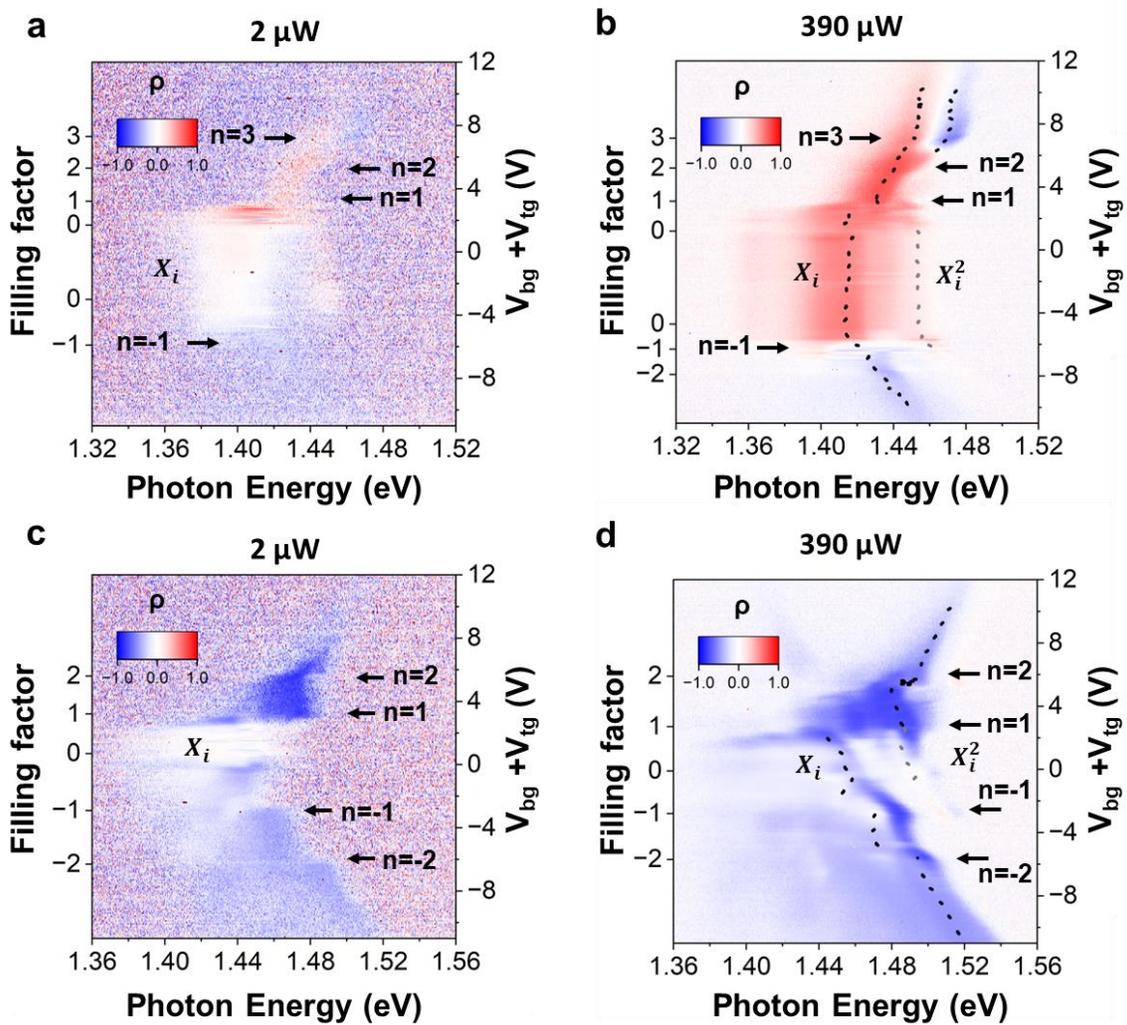

**Fig.4 | Valley polarization of the Excitonic Mott insulator.** (a) and (b) are doping-dependent circular polarization of PL of the R-stacked device R1 under an excitation power of 2 μW (average exciton density ~ $3.5 \times 10^{11}$ cm$^{-2}$) and 390 μW (average exciton density ~ $6.9 \times 10^{13}$ cm$^{-2}$), respectively. (c) and (d) are doping-dependent circular polarization of PL of the H-stacked device H1 under an excitation power of 2 μW and 390 μW, respectively. A CW laser with photon energy centered at 1.70 eV was used for the optical excitation, and all data were taken at a temperature of 5.0 K.

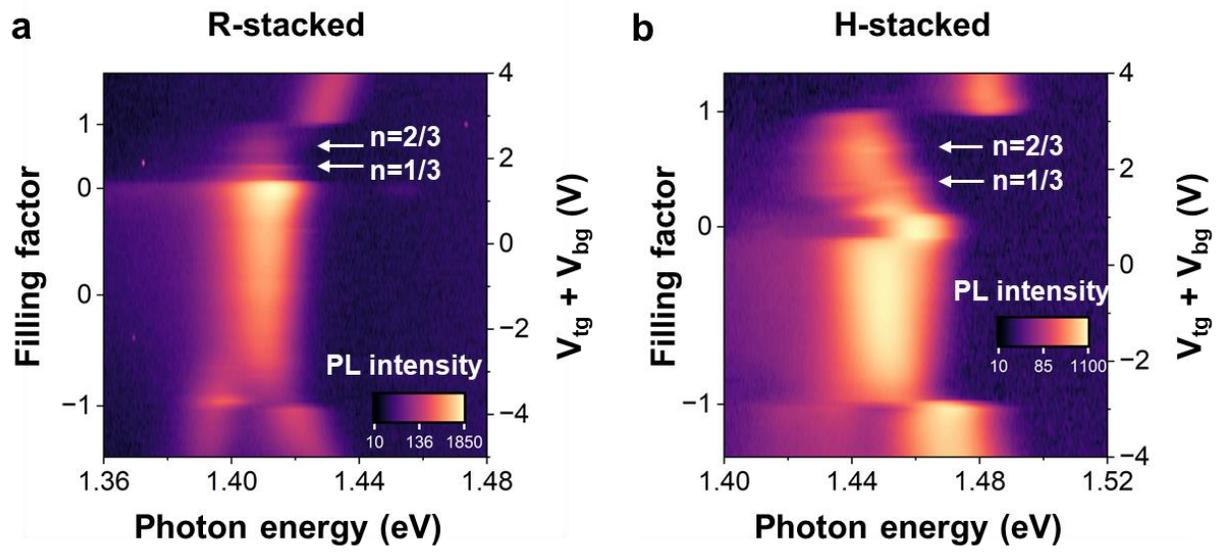

**Extended Data Fig.1 | Zoom-ins of the doping-dependent PL spectra.** (a) and (b) are the Zoom-ins of doping-dependent PL spectra in Figs. 1 c, d, respectively.

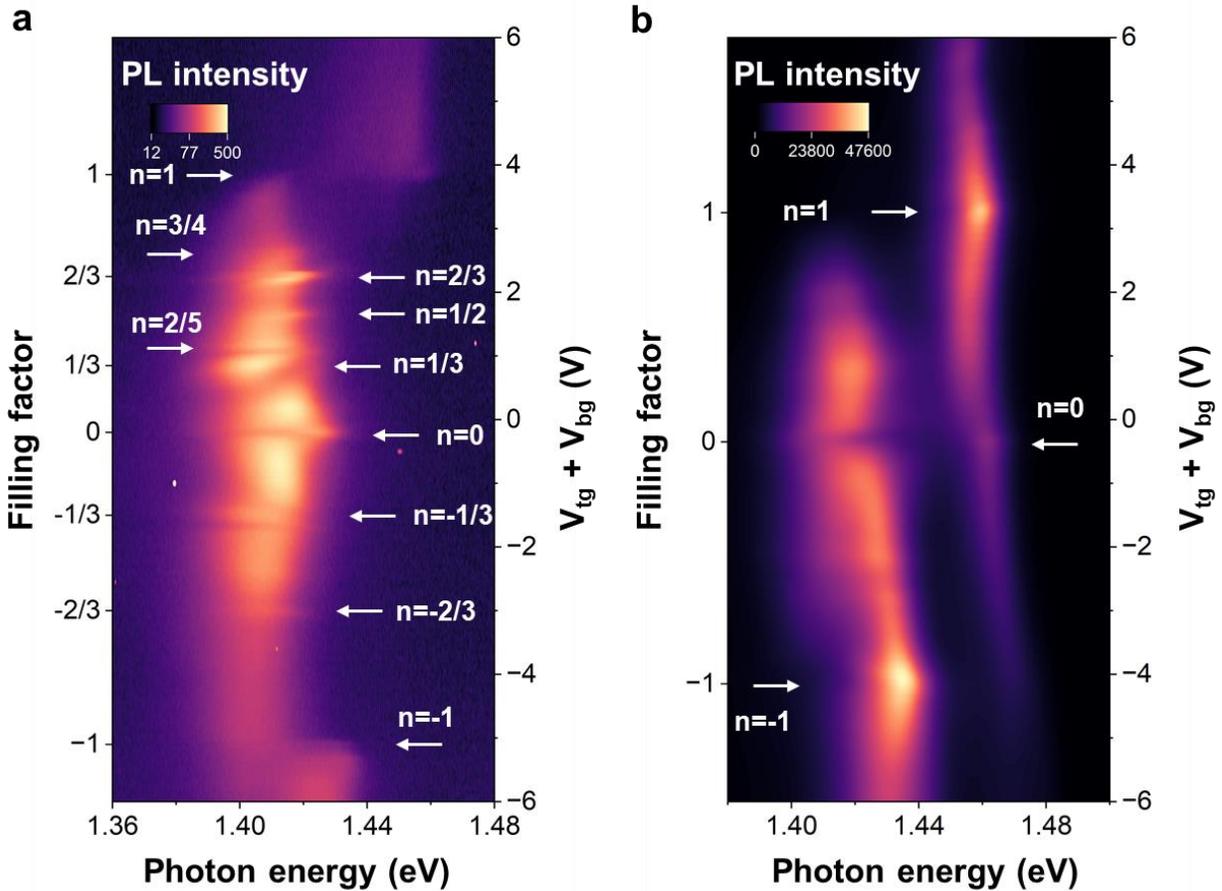

**Extended Data Fig.2 | PL spectra of an H-stacked device showing more fractional fillings.** (a) and (b) doping-dependent PL spectra measured on device H5 using an excitation power of 0.2 µW and 300 µW, respectively. A CW laser with photon energy centered at 1.70 eV was used for the optical excitation, and all data were taken at a temperature of 5.0 K.

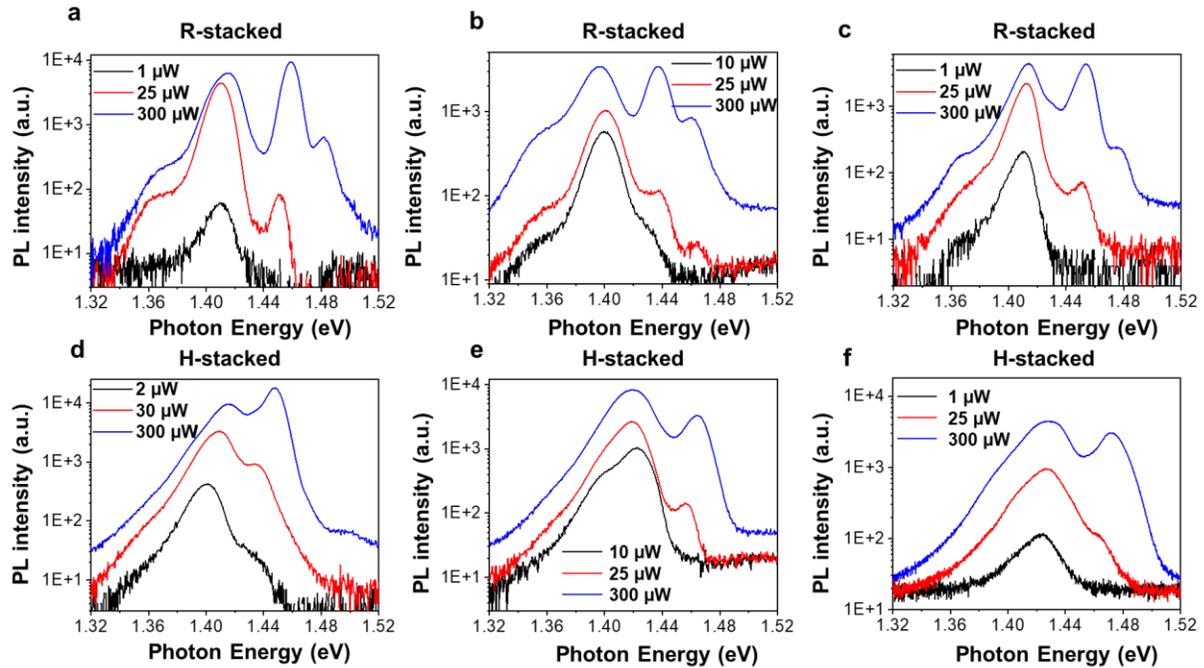

**Extended Data Fig.3 | Power-dependent PL spectra for different R-stacked and H-stacked devices around n=0**. (a), (b) and (c) are power-dependent PL spectra of R-stacked devices R1, R2 and R3, respectively. (d), (e) and (f) are power-dependent PL spectra of H-stacked devices H2, H3 and H4, respectively. Black, red, and blue line show PL spectra under low, medium, and high excitation power. A CW laser with photon energy centered at 1.96 eV was used for the optical excitation, and all data were taken at a temperature of 4-10 K.

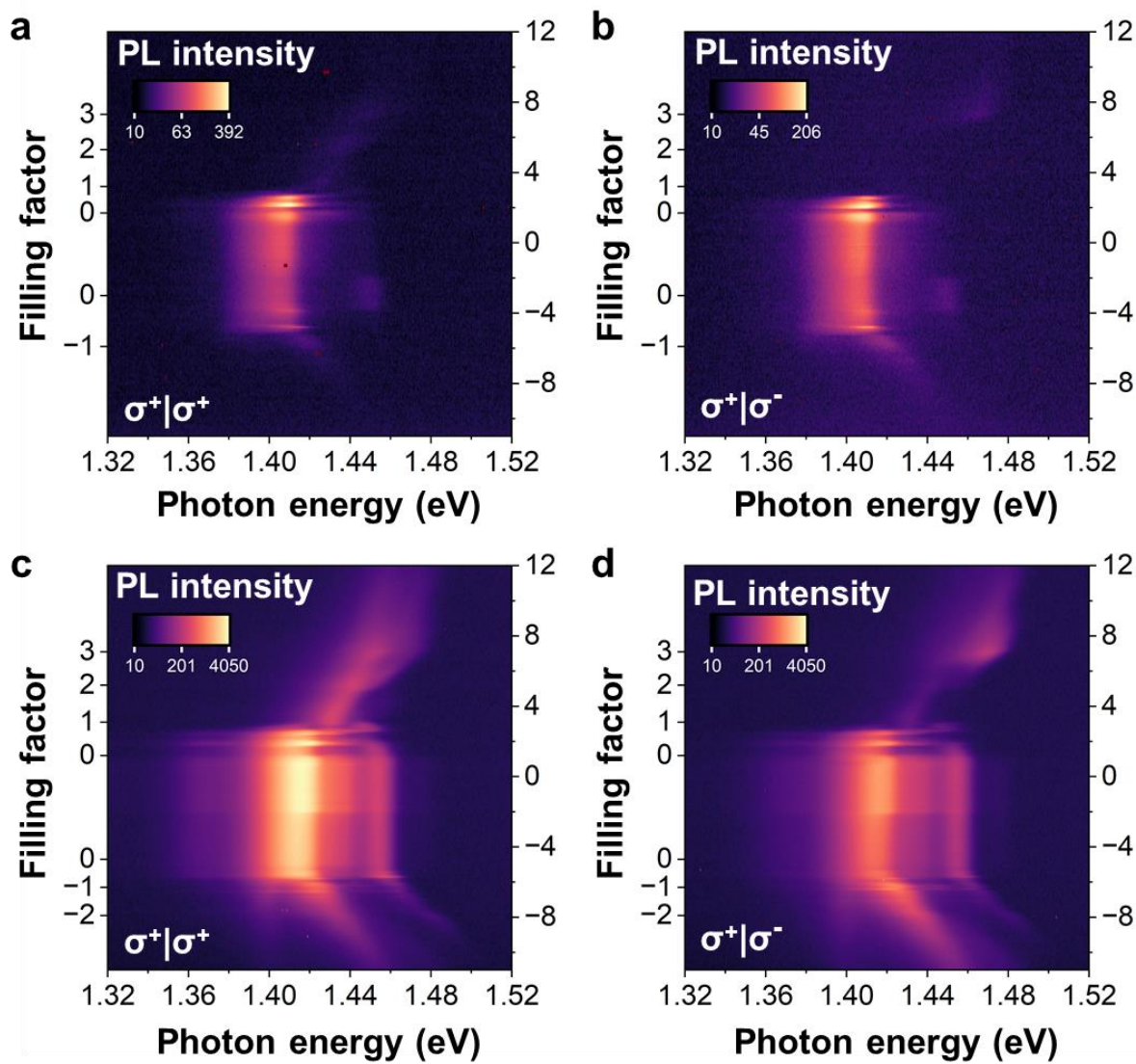

**Extended Data Fig.4 | Helicity-resolved doping-dependent PL spectra for R-stacked WS$_2$/WSe$_2$ moiré bilayers under different excitation powers.** (a) and (b) are measured right ($\sigma^+$) and left circularly ($\sigma^-$) polarized doping-dependent PL spectra for R1, under a $\sigma^+$ excitation with the excitation power of 2 µW. (c) and (d) are $\sigma^+$ and $\sigma^-$ PL spectra under a $\sigma^+$ excitation with an excitation power of 390 µW. A CW laser with photon energy centered at 1.70 eV was used for the optical excitation, and all data were taken at a temperature of 5.0 K.

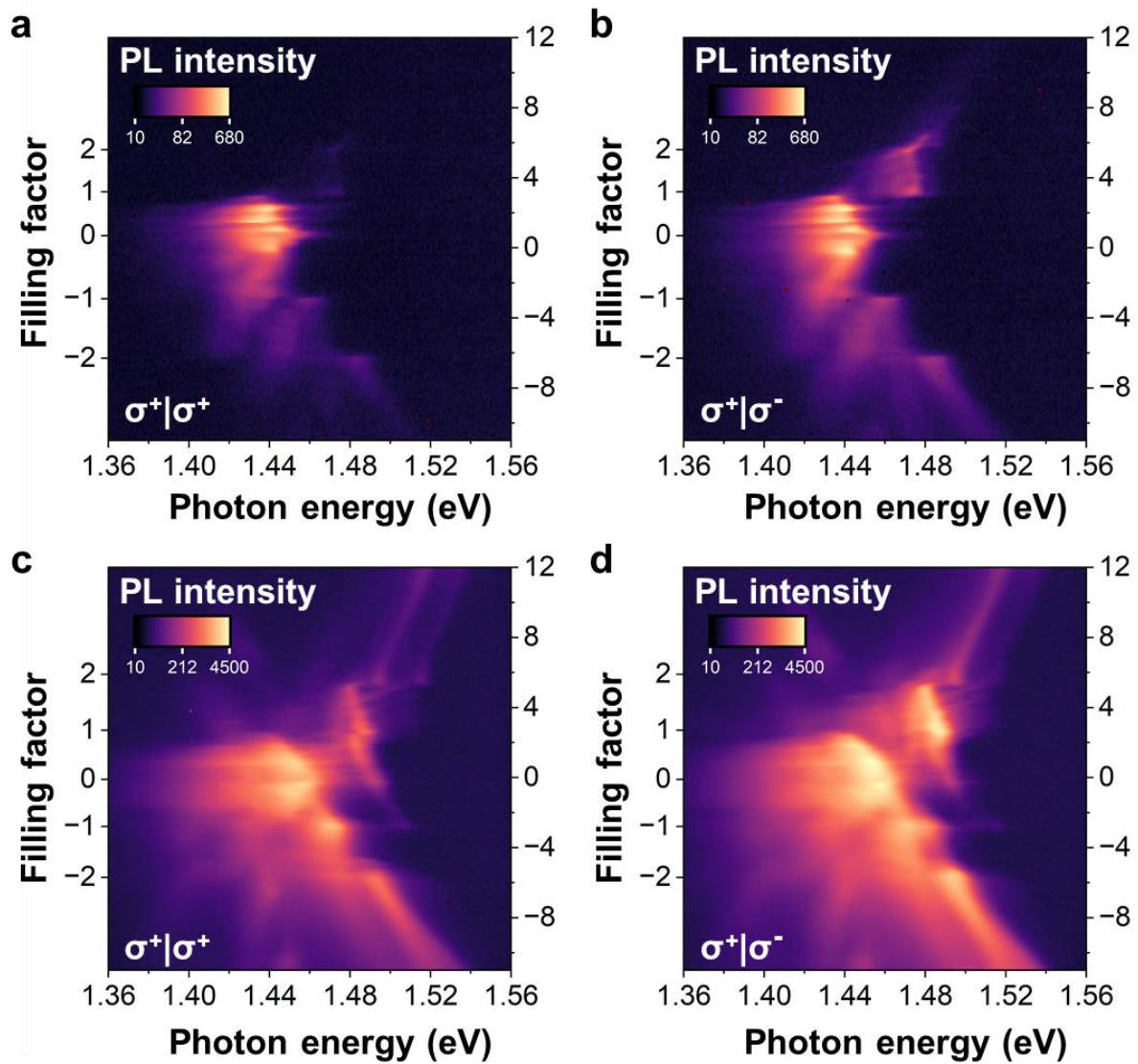

**Extended Data Fig.5 | Helicity-resolved doping-dependent PL spectra for H-stacked moiré bilayers under different power.** (a) and (b) are measured right (σ⁺) and left circularly (σ⁻) polarized doping-dependent PL spectra for device H1, under a σ⁺ excitation with an excitation power of 2 µW. (c) and (d) are σ⁺ and σ⁻ PL spectra under a σ⁺ excitation with an excitation power of 390 µW. A CW laser with photon energy centered at 1.70 eV was used for the optical excitation, and all data were taken at a temperature of 5.0 K.

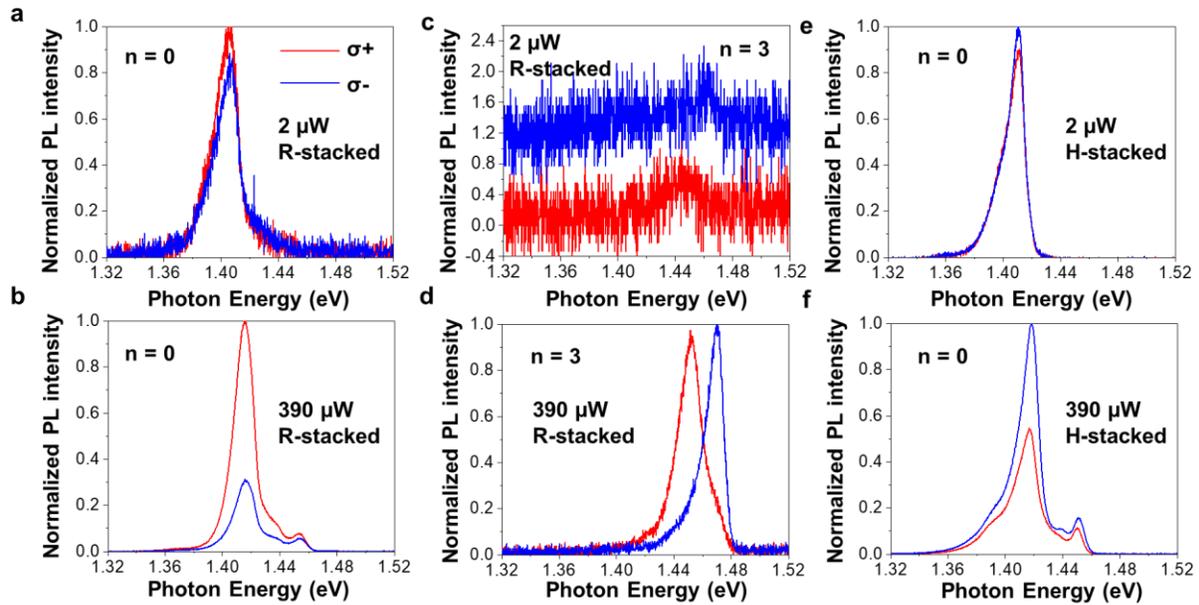

**Extended Data Fig.6 | Helicity-resolved PL spectra of device R1 and H1 for various filling factors and under different excitation powers**. (a) and (b) are PL spectra of R1 at n=0 under the excitation power of 2 µW and 390 µW, respectively. The red and blue lines represent measured right (σ⁺) and left circularly (σ⁻) PL spectra, respectively, under the σ⁺ excitation. (c) and (d) are PL spectra of R1 at n=3 under the excitation power of 2 µW and 390 µW, respectively. (e) and (f) are PL spectra of H1 at n=0 with excitation power of 2 µW and 390 µW, respectively. These data are extracted from Extended Data Figs. 4 and 5. The data in (c) are offset for the clarity of presentation.